\documentclass[12pt]{article}
\usepackage{amsmath}
\usepackage{amssymb}
\usepackage{amsfonts}
\usepackage{graphicx}

\newcommand{\ov } {\over }
\newcommand{\be}{\begin{equation}}
\newcommand{\ee}{\end{equation}}
\newcommand{\bea}{\begin{eqnarray}}
\newcommand{\eea}{\end{eqnarray}}

\begin{document}

\bigskip%
\begin{titlepage}
\begin{flushright}
UUITP-06/08\\
\end{flushright}
\vspace{1cm}
\begin{center}
{\Large\bf Chain inflation revisited\\}
\end{center}
\vspace{3mm}
\begin{center}
{\large
Diego Chialva$^1$ and Ulf H.\ Danielsson$^2$} \\
\vspace{5mm}
Institutionen f\"or fysik och astronomi \\
Uppsala Universitet,
Box 803, SE-751 08
Uppsala, Sweden

\vspace{5mm}

{\tt
{$^1$}diego.chialva@fysast.uu.se \qquad
{$^2$}ulf.danielsson@fysast.uu.se \\
}
\end{center}
\vspace{5mm}

\begin{center}
{\large \bf Abstract}
\end{center}

\noindent
This paper represents an in-depth treatment of the
chain inflation scenario. We fully determine the evolution of the universe in
the model, the necessary conditions
in order to have a successful inflationary period, and the
matching with the observational results regarding the 
cosmological perturbations. We study in great detail, and in general,
the dynamics of the background, as well as
the
mechanism of generation of the perturbations. We also find an explicit
formula for the spectrum of adiabatic perturbations. Our results prove that
chain inflation is a viable model for solving the horizon, entropy and
flatness problem of standard cosmology and for generating the right
amount of adiabatic cosmological perturbations. The results are radically
different from those found in previous works on the subject.
Finally, we argue that there is a natural way to embed chain inflation
into flux compactified string theory. We discuss 
the details of the implementation and how to fit observations.

\vfill
\begin{flushleft}
April 2008
\end{flushleft}
\end{titlepage}\newpage

\tableofcontents

\section{Introduction}

String theory is our best developed theory of quantum gravity and should
describe the evolution and birth of our universe. In particular, it should
provide a mechanism for inflation. Unfortunately, all models constructed so
far suffer from various deficiencies. A significant difficulty is that
potentials appropriate for slow-roll seem to be uncommon in string theory.
On the other hand, the theory is in many cases poorly known, and therefore
so are the corrections to the potentials, which can be important for large
values of the fields.

Most successful models of inflation are based on second order phase
transitions with a field that rolls slowly through field configuration
space. In the simplest models the field travels long distances of the order
of the Planck mass, which makes a low energy approximation questionable.
There is also the issue of fine-tuning of the models if one wants to obtain
enough inflation.

In this paper we will explore the possibility of using a series of first
order phase transitions to generate inflation. While it is well known that a
single first order phase transition, as in the case of old inflation, cannot
do the job, the situation could be dramatically different with a large
number of transitions. The idea is called \textit{chain inflation} \cite%
{Freese:2004vs, Freese:2006fk}. The basic idea is that inflation need not
proceed in one step. Instead there is a long series of metastable minima
where the inflaton field or fields spend just a short time allowing for a
fraction of an e-folding. The field then tunnels to the next metastable
minima in the series. Only after adding up several such steps will there be
a sufficient number of e-foldings.

It is widely believed that string theory has a rich structure of many
metastable vacua separated by potential barriers related to domain walls (D-
and M-branes). Such a setup favours first order transitions, and should
provide suitable conditions for chain inflation. The basic idea is that the
early universe searched its way through the landscape experiencing a large
number of tunneling events before settling down in the present state. During
this process the universe should have gone through periods of inflation.

Our motivation for studying chain inflation is twofold. On the one hand,
this scenario has not been studied in sufficient detail to make it robust.
Moreover, results in previous works are not fully consistent with a complete
treatment of the subject, \cite{Freese:2004vs, Freese:2006fk, Freese:2005kt,
Feldstein:2006hm, Watson:2006px, Huang:2007ek, Huang:2008jr}. On the other
hand, our interest is directly related to a previous study of the topography
of the string landscape. In \cite{DanJohLar, ChDaJoLaVo} it was discovered
that series of simply related vacua are a generic feature of the string
landscape in the context of flux compactifications (for a review see for
example \cite{Douglas:2006es, Grana:2005jc} and references therein).
Contrary to previous studies of chain inflation, this provides us with a
reasonably constrained framework for implementing chain inflation.
Furthermore, in this case we are able to propose an inflationary model where
the agents are complex structure moduli, in contrast to the models so far
described in the literature, which involve K\"{a}hler moduli or open string
moduli.

The paper is divided into four main sections. In section \ref{secbackgen} we
discuss the basic problems of inflation with first order phase transitions,
and discuss the various constraints that need to be met for successfully
solving the horizon, entropy and flatness problem and achieving percolation
and large scale thermalization. In section \ref{secpert} we discuss the
generation of cosmological perturbations in chain inflation, proposing a
specific set of equations for determining them and computing the power
spectrum.

In the first sections our discussion is at a general level, while a detailed
model based on string theory is specified in section \ref{secmodel}. There,
we review the construction of a series of minima using flux compactification
and apply it to chain inflation. We compare our results to the experimental
data in section \ref{matchdata}. We then end with some conclusions,
speculations and outlook.

\bigskip


\section{Successful chain inflation}

\label{secbackgen} 

\bigskip

The key-point of chain inflation is that the slow-rolling of the inflaton is
substituted by a series of subsequent tunneling processes. Chain inflation
must still fulfill the same demands as ordinary inflation with a slow
evolution of the energy density of the vacuum allowing for an accelerated
expansion of the universe. At the same time, in order to obtain a
homogeneous and isotropic universe, we need the tunneling events to proceed
in such a way as to guarantee percolation and homogenization at every step
of the ladder of transitions.

While these requirements are enough\footnote{%
A viable model for  inflation must also provide a suitable mechanism for
reheating, but we will not address this question in this paper.} to solve
the horizon, entropy and flatness problems of standard cosmology (we can
call it \emph{background requirements}), the model must also provide the
right amount of density perturbations and the correct features of the
spectrum.

We will deal with this last aspect later in the paper (section \ref{secpert}%
), for the moment let us proceed by establishing the background requirements
for the tunneling processes.

\subsection{Basic setup}

Consider a theory with a potential with many metastable minima at different
energies\footnote{%
If we are allowed to consider the string landscape as governed by a very
complicated effective action with a potential depending on many fields, this
description applies to that case as well.}. The number of fields could in
principle be very large, as it is supposed to be for string moduli, and
their dynamics as they pass through the various minima could be complex,
with rolling, tunneling, and jumping phases. In the following we will
simplify the description by modeling the dynamics purely through the stress
energy tensors of the fields and the gravity field.

The phase transitions among the various vacua proceed by nucleation of
bubbles. The bubbles are filled by the new vacuum at a lower energy, and all
the latent heat of the transition is stored within the walls of the bubbles.
When bubbles collide the energy of the walls is converted in radiation
(massless string states). The process of bubble collision and the transfer
of the energy to radiation are complex phenomena, that can be studied
numerically. We are not going to do so, and will instead stick to a
simplified model based on a number of assumptions:

\begin{itemize}
\item we consider values for the decay rate per unit physical volume $\Gamma
_{n}$ between vacua $n$ and $n-1$ such that percolation and large scale
thermalization is achieved at every phase transition (we will investigate
which values allow this in the following)

\item decays between two distant vacua are strongly suppressed

\item the bubbles are produced with negligible volume and expand at the
speed of light

\item the energy of the walls is optimally transferred to radiation in the
collision and the thermalization time is very short
\end{itemize}

In this way, we can assume that a homogeneous and isotropic space-time is
obtained in a very short time in every patch and phase, and we can therefore
describe the metric using the Friedman-Robertson formula 
\begin{equation}
ds^{2}=-dt^{2}+a(t)^{2}d{\mathbf{x}}^{2}.
\end{equation}

Schematically we have the following picture. As time passes the universe is
divided into regions of different phases (new vacua) created through bubble
nucleation and rapid percolation. The collisions among bubble walls produce
an amount of radiation which remains trapped within every phase. Our actual
Universe is dominated by one of these phases. From our point of view,
therefore, we observe the Universe as filled with the radiation produced
from the collisions of the walls of the bubbles of our phase and small
patches, distributed in an homogeneous way, composed of the previous phases
(which are contained one within the other, together with the radiation
produced at every step).

It is convenient to write the equations involving the energy tensors of the
vacuum and the radiation by formally distinguishing every phase with an
index $m$ and associating to each one a vacuum and a radiation stress-energy
tensor, distinguished by $\ell =V,r$ 
\begin{equation}
T_{\mu \nu }^{\ell ,m}=(\rho ^{\ell ,m}+p^{\ell ,m})u_{\mu }^{\ell ,m}u_{\nu
}^{\ell ,m}+p^{\ell ,m}g_{\mu \nu },
\end{equation}%
where $u_{\mu }^{\ell ,m},\,\rho ^{\ell ,m},\,P^{\ell ,m},$ for $\ell
=\{r,V\},$ are the proper four-velocity, energy and pressure densities. We
write the energy density, pressure and stress-energy tensors for radiation
and vacuum as 
\begin{equation}
\rho ^{\ell }\equiv \sum_{m}\rho ^{\ell ,m}\qquad P^{\ell }\equiv
\sum_{m}P^{\ell ,m}\qquad T_{\mu \nu }^{\ell }\equiv \sum_{m}T_{\mu \nu
}^{\ell ,m},  \label{backlversmpart}
\end{equation}%
and the total energy density, pressure and stress-energy tensor as 
\begin{equation}
\rho \equiv \sum_{\ell }\rho ^{\ell }\qquad P\equiv \sum_{\ell }P^{\ell
}\qquad T_{\mu \nu }\equiv \sum_{\ell }T_{\mu \nu }^{\ell }.
\label{backtotverspart}
\end{equation}%
We furthermore write 
\begin{equation}
\rho ^{V}\equiv \rho ^{\mathcal{V}}+\rho ^{W},  \label{whatrhoV}
\end{equation}%
expressing that the \textquotedblleft vacuum energy\textquotedblright\ is
actually the sum of the energy density in the interior of the bubbles, $\rho
^{\mathcal{V}}$, plus the energy stored in the bubbles walls, $\rho ^{W}$.

The energy and pressure densities of our generalized fluids\footnote{%
As well known, the hydrodynamical fluid description do not apply for scalar
fields and dark energy, and therefore we cannot use it for modeling the
dynamics of our vacuum fields here. Nevertheless, it is possible and
well-defined to speak of energy density and radiation with reference to the
components of the energy tensor, that is why we speak of generalized fluids.}
depend on space and time, but, due to the hypothesis of homogeneity and
isotropy within every phase, their average over the volume occupied by the
phases depends only on time 
\begin{equation}
\langle \rho ^{V,r}({\mathbf{x}},t)\rangle =\rho ^{V,r}(t).
\end{equation}%
Written in terms of $\rho ^{V}$ and $\rho ^{r}$, the Friedman and Chauduri
equations are 
\begin{eqnarray}
H^{2} &=&\frac{8\pi G}{3}(\rho ^{V}+\rho ^{r})  \label{backFried} \\
\dot{H} &=&-4\pi G\sum_{\ell =V,r}\left( \rho ^{\ell }+P^{\ell }\right) .
\label{backChaud}
\end{eqnarray}

\subsection{A slow roll through the landscape}

\bigskip

In order to have a successful chain inflation mechanism, the system must
undergo a series of successive tunnelings such that the decay of the vacuum
energy is effectively slowed down. In this way inflation can effectively
cure the problems related to entropy, flatness and horizon size, by lasting
for a sufficient number of e-foldings ($N_{e}\sim 60$).

To avoid a too rapid decay of the vacuum energy it is necessary that each
tunneling step occurs only after a suitable and sufficient fraction of the
universe volume has completed the preceding tunneling event. We will show
how this condition arises through multiple tunnelings, and thereby establish
that chain inflation is a viable model within string theory or other gravity
theories with multiple metastable vacua. Although our results are in
agreement with the chain inflation models expectations, our formulas for the
energy density of the vacua, their lifetime and persistence are in fact very
different from those usually found in the literature (see for instance \cite%
{Freese:2004vs, Freese:2006fk, Freese:2005kt, Feldstein:2006hm,
Watson:2006px, Huang:2007ek, Huang:2008jr}). In many cases the results have
not been derived, but instead inferred from calculations based on Old
Inflation, not acknowledging that in that case only one tunneling step is
considered.

To proceed we write the various contributions to the energy density as 
\begin{eqnarray}
\rho ^{\mathcal{V}}(t) &=&\sum_{m=0}\epsilon _{m}p_{m}(t)  \label{endensav3}
\\
\rho ^{W}(t) &=&\sum_{m=1}\Delta \epsilon _{m}\sum_{n=0}^{m-1}p_{n}(t)%
\mathcal{F}_{m,m-1}(t)  \label{endensawall3} \\
\dot{\rho}^{r}(t) &=&-4H\rho ^{r}+\sum_{m=1}\Delta \epsilon _{m}\partial
_{t}\left( 1-\sum_{n=0}^{m-1}p_{n}(t)\mathcal{F}_{m,m-1}(t)\right) ,
\label{endensarad3}
\end{eqnarray}%
where $\epsilon _{m}$ is the value of the potential of the potential
energy\footnote{Note that in general $\epsilon _{0}\neq 0$.}
in the vacuum $m$, $\Delta \epsilon _{m}\equiv (\epsilon _{m}-\epsilon
_{m-1})$, \,\,$p_{m}(t)$ is the fraction of volume still occupied by the
vacuum $m$ at time $t$ and, finally, $\mathcal{F}_{m,m-1}(t)$ is the energy
weighted fraction of uncollided walls at time $t$.
The energy density of the $m$%
-th vacuum phase is therefore given by 
\begin{equation}
\rho ^{V,m}= \epsilon _{m}p_{m}+\Delta \epsilon _{m+1}\sum_{n=0}^{m}p_{n}(t)%
\mathcal{F}_{m+1,m}(t).  \label{rhoVm}
\end{equation}

In chain inflation the $p_{n}(t)$ are obtained from a coupled set of
differential equations:%
\begin{eqnarray}
\dot{p}_{N} &=&-\widetilde{\Gamma }_{N}p_{N}  \notag  \label{systemprob} \\
\dot{p}_{N-1} &=&-\widetilde{\Gamma }_{N-1}p_{N-1}+\widetilde{\Gamma }%
_{N}p_{N}  \notag \\
&&... \\
\dot{p}_{0} &=&\widetilde{\Gamma }_{1}p_{1},  \notag
\end{eqnarray}%
with%
\begin{equation*}
\sum_{m}p_{m}(t)=1.
\end{equation*}%
We have 
\begin{equation}
\widetilde{\Gamma }_{m}(t)=\Gamma _{m}V^{\text{physical}}(t_{0},t),
\label{reltildenakedgamma}
\end{equation}%
where $\Gamma _{m}$ is the decay rate per unit time and (physical) volume.
As we will see later, the amount of radiation is negligible and therefore
the decay rate will essentially be given by the Coleman-Callan tunneling
process, and not by thermal nucleation of bubbles.

The uncollided fraction of walls of bubbles of the phase $m-1$ is given by
(see also \cite{TurWeiWid}) 
\begin{equation}
\mathcal{F}_{m,m-1}(t)={\int_{0}^{t}dt^{\prime }\Gamma _{m}V^{\text{physical}%
}(t,t^{\prime })p_{m}(t^{\prime })p_{m,\text{un}}(t,t^{\prime })\ov%
\int_{0}^{t}dt^{\prime }\Gamma _{m}V^{\text{physical}}(t,t^{\prime
})p_{m}(t^{\prime })},  \label{fracuncoll}
\end{equation}%
where $p_{m,\text{un}}(t,t^{\prime })$ is the fraction of a (reference) wall
generated at time $t^{\prime }$ which is still not collided at time $t$.

This fraction represents the probability that a point of the (reference)
wall has not collided at time $t$, and is equal to the probability that a
point just outside the wall is still in the phase $m$, so that: 
\begin{equation}
p_{m,\text{un}}(t,t^{\prime })=e^{-\left( \int_{0}^{t}\widetilde{\Gamma}%
_m-\int_{0}^{t^{\prime }}\widetilde{\Gamma}_m\right) }.  \label{probuncoll}
\end{equation}%
Note that $p_{m,\text{un}}(t^{\prime },t^{\prime })=1$, which simply means
that at the time of nucleation the wall is completely uncollided.

The solution of the system (\ref{systemprob}) is different depending on
whether the nucleation rates per time $\widetilde{\Gamma }_{m}$ are

\begin{itemize}
\item[(\textbf{a})] equal for every $m$ (that is $\widetilde{\Gamma }_{m}=%
\widetilde{\Gamma }$)

\item[(\textbf{b})] different for every $m$
\end{itemize}

We will show how the correct requirement for having a successful
inflationary periods can arise, so that chain inflation is a viable model
within string theory or other gravity theories with multiple metastable
vacua. Scenario (b) will be dealt with in detail in section \ref{difftunn},
and we concentrate on case (a) in the following section \ref{eqtunn}.

\subsubsection{Percolation and large scale thermalization}

We must now discuss for which values of the parameters in our model
(essentially the decay rates) we are able to achieve percolation (and large
scale thermalization) at every phase transition. There are three time scales
that can be used to study the final stage of an inflationary expansion,\cite%
{TurWeiWid, GutWeinRecov}.

\begin{itemize}
\item $\mathbf{t_{c}^{m}}$: the time when the volume of the patch starts to
contract, that is the time when 
\begin{equation}
V_{phys,m}^{-1}{dV_{phys,m}\ov dt}
\end{equation}%
becomes negative.

By using $V_{phys,m}(t)=a^{3}(t)p_{m}(t),$ it is easily seen that $t_{c}^{m}$
is given by the formula 
\begin{equation}
3H(t_{c}^{m})+{\dot{p}_{m}\ov p_{m}}|_{t=t_{c}^{m}}\leq 0\quad \Rightarrow
\quad \widetilde{\Gamma }_{m+1}{p_{m+1}(t_{c}^{m})\ov p_{m}(t_{c}^{m})}\leq%
\widetilde{\Gamma }_{m}-3H(t_{c}^{m})  \label{deftcm}
\end{equation}%
and, in particular $\widetilde{\Gamma }_{m}-3H(t_{c}^{m})\geq 0$.

Note that (\ref{deftcm}) implies $\dot{p}_{m}|_{t=t_{c}^{m}}<0,$and
therefore $t_{c}^{m}>t_{\text{max}}^{m},$ where $t_{\text{max}}^{m}$ is the
time for which $p_{m}$ is at its maximum (given by $\dot{p}_{m}|_{t=t_{\text{%
max}}^{m}}=0$).

\vskip0.2cm

\item $\mathbf{t_{p}^{m}}$: the percolation time for the $m$-th phase, given
by the condition 
\begin{equation}
{\text{total volume of bubbles of phase $m$}\ov\text{total volume}}%
|_{t_{p}^{m}}\gtrsim 0.34
\end{equation}

$t_{p}^{m}$ can be found in the following way,\cite{GutWeinRecov}. The
bubble nucleation rate per unit time and coordinate volume is 
\begin{equation}
{d\Upsilon \ov dtd^{3}x}=a^{3}(t)\,\Gamma _{m+1}\,\theta (t-t_{B}),
\end{equation}%
where $t_{B}$ represents the time when the nucleation of bubbles starts.

Since the bubbles expand at the speed of light (in our model) we can relate
their radius to the time using the formula $dr=a(t)^{-1}dt$. We choose a
reference radius $r_{1}$, corresponding to a time $t_{r_{1}}$. Bubbles with
radius $r>r_{1}$ would nucleate at times $t<t_{r_{1}}$. Therefore the
nucleation rate per unit coordinate volume for a sphere having $r>r_{1}$ is
given by 
\begin{equation}
{d\Upsilon \ov d^{3}x}|_{r>r_{1}}=\Gamma_{m+1}%
\int_{t_{B}}^{t_{r_{1}}}a^{3}(t)dt.
\end{equation}%
Successful percolation (and, as can be shown, also large scale
thermalization, see \cite{GutWeinRecov}) at time $t_{p}^{m}$, for $r_{1} $
small, can then be achieved if 
\begin{equation}
{\text{total volume of bubbles with $r>r_{1}$}\ov\text{total volume}}%
|_{t_{p}^{m}}>{4\pi \ov3}r_{1}^{3}{d\Upsilon (t_{p}^{m})\ov d^{3}x}%
|_{r>r_{1}}\gtrsim 0.34,  \label{deftpm}
\end{equation}%
where ${4\pi \ov3}r_{1}^{3}$ is the volume of bubbles of radius $r_{1}$.

\vskip0.2cm

\item $\mathbf{t_{sf}^{m}}$: the time by which the patch of phase $m$ has
become a small fraction of the total volume, which for example can be chosen
to correspond to the moment when $p_{m}(t_{sf}^{m})=0.01$.
\end{itemize}

We will analyze these requirements in more details in the following
sections, when we will have the explicit form of $p_m(t)$.

\bigskip

\subsection{Equal tunneling rate at each step of the ladder}

\label{eqtunn}

\bigskip

We will now assume that the tunneling rates are constant and equal at every
step of the ladder of decay events. That is 
\begin{equation}
\widetilde{\Gamma }_{m}(t)\sim \widetilde{\Gamma }.
\end{equation}%
In this case the solution to (\ref{systemprob}) is easily seen to be%
\begin{equation}
p_{N-k}(t)=\frac{\left( \tilde{\Gamma}t\right) ^{k}}{k!}e^{-\tilde{\Gamma}t}.
\label{prosoleqdec}
\end{equation}%
Furthermore, the uncollided fraction of walls of bubbles of the phase $N-k-1$
is given by 
\begin{equation}
\mathcal{F}_{N-k,N-k-1}(t)=\frac{{p_{N-k-1}(t)\,k!}}{{k!-\Gamma (k+1,t)}},
\end{equation}%
where 
\begin{equation}
{\Gamma (k+1,t)}={k!}e^{-y}\sum_{s=0}^{k}\frac{y^{s}}{s!},
\end{equation}%
is the incomplete Gamma function.

It is now possible to determine for which values of the decay rate $%
\widetilde{\Gamma }$ successful percolation and large scale thermalization
are achieved. Formula (\ref{deftcm}) for $m=N-k$ now reads 
\begin{equation}
t_{c}^{N-k}\geq {k\ov\widetilde{\Gamma }-3H,}
\end{equation}%
and if we take $H$ to be constant during inflation\footnote{%
The accuracy of this approximation can be verified by solving the Friedman
and Chauduri equations using the formula for the energy density of the
vacuum that we will provide in (\ref{rhovacacc}).}, we see from formula (\ref%
{deftpm}) that 
\begin{equation}
{\widetilde{\Gamma }\ov3H}(1-e^{3H(t_{r_{1}}^{m}-t_{p}^{m})})>0.34\quad
\Rightarrow \quad t_{p}^{m}>-{\log (1-0.34{3H\ov\widetilde{\Gamma }})\ov3H,}
\end{equation}%
where we have used (\ref{reltildenakedgamma}) to rewrite the formula in
terms of $\widetilde{\Gamma }$ and we have sent $t_{r_{1}}\rightarrow 0$. We
can see that if $\widetilde{\Gamma }$ is sufficiently larger than $3H$, then
the contraction of volume, percolation and large scale thermalization are
achieved in very short times.

We can now calculate the vacuum energy, which is given by (see (\ref{rhoVm}%
)) 
\begin{eqnarray}
\rho ^{V}(t) &=&\rho ^{\mathcal{V}}(t)+\rho ^{W}(t)  \notag \\
&=&e^{-y}\sum_{k=1}^{N}\epsilon _{N-k}\frac{y^{k}}{k!}\left( 1+{\Delta
\epsilon _{N-k+1}\ov\epsilon _{N-k}}\frac{\Gamma (N+1, y)(k-1)!-\Gamma (k,
y)N!}{N!((k-1)!-\Gamma (k,y))}\right) +\epsilon _{N}e^{-y},  \notag \\
\end{eqnarray}%
where we have used the variable $y=\widetilde{\Gamma }t\equiv {t \ov
\tau}$.

Apart from the last steps of tunneling, it is possible to see that we have  
\begin{equation}
{\rho ^{\mathcal{V},N-k} \ov \epsilon_{N-k}} \approx {\rho ^{W,N-k} \ov %
\Delta \epsilon _{N-k+1}}.  \label{wallneglect}
\end{equation}%
Note, then, that the transfer of energy from the vacuum to radiation is
quite low during the main part of the tunneling chain. We can then write:  
\begin{equation}
\rho ^{V}(t) \sim e^{-y}\sum_{k=1}^{N}\epsilon _{N-k+1}\frac{y^{k}}{k!}%
+\epsilon _{N}e^{-y},
\end{equation}

Motivated by the specific model we will describe in section \ref{secmodel},
we consider two possible functional forms for the energy density at every
stage 
\begin{equation}  \label{modelsenergy}
\epsilon _{n}=%
\begin{cases}
m_{f}^{4}n & \text{case I} \\ 
{m_{f}^{4}\ov2}n^{2} & \text{case II}%
\end{cases}%
,
\end{equation}%
where $m_{f}$ is a suitable energy scale. In this way we can easily compute
the vacuum energy with the result 
\begin{equation}
\rho ^{V}(t)\sim%
\begin{cases}
\frac{m_{f}^{4}}{N!}\left( \left( N+1-\frac{t}{\tau }\right) \Gamma \left(
N+1,\frac{t}{\tau }\right) +\left( \frac{t}{\tau }\right) ^{N+1}e^{-\frac{t}{%
\tau }}\right) & \text{case I} \\ 
\frac{m_{f}^{4}}{2N!}\left( \left( \left( N+1-\frac{t}{\tau }\right) ^{2}+%
\frac{t}{\tau }\right) \Gamma \left( N+1,\frac{t}{\tau }\right) +\left( N-%
\frac{t}{\tau }+1\right) \left( \frac{t}{\tau }\right) ^{N+1}e^{-\frac{t}{%
\tau }}\right) & \text{case II}%
\end{cases}%
.  \label{vacenerfor}
\end{equation}%
If $N$ is large, the incomplete gamma function is to a very high degree of
approximation constant, $\Gamma \left( N+1,\frac{t}{\tau }\right) \sim N!,$
for values of $t$ almost all the way up to $N\tau $. We also have that $%
\left( \frac{t}{\tau }\right) ^{N}e^{-\frac{t}{\tau }}\ll \Gamma \left( N+1,%
\frac{t}{\tau }\right) $. As a consequence we find, to a high degree of
accuracy, 
\begin{equation}  \label{rhovacacc}
\rho ^{V}(t)\sim 
\begin{cases}
m_{f}^{4}\left( N+1-\frac{t}{\tau }\right) & \text{case I} \\ 
{m_{f}^{4}\ov2}\left( N+1-\frac{t}{\tau }\right) ^{2} & \text{case II,}%
\end{cases}%
\end{equation}%
which indeed leads to the expected slow roll.

It is convenient to define an effective slow-roll parameter 
\begin{equation}
\varepsilon =-\frac{{\dot{H}}}{{H^2}}.
\end{equation}%
Using the excellent approximation%
\begin{equation}
n=N+1-\frac{t}{\tau },
\end{equation}%
it can be written 
\begin{equation}
\varepsilon \equiv -\frac{\dot{H}}{H^{2}}=%
\begin{cases}
\frac{1}{2nH\tau } & \text{case I} \\ 
\frac{1}{nH\tau } & \text{case II}%
\end{cases}%
.  \label{slowrollpar}
\end{equation}

For what concerns the number of e-foldings ($N_{e}$), a
straightforward computation gives 
\begin{equation}
 N_{e}=\int Hdt\sim 
\begin{cases}
{1\ov3\varepsilon }{n_{\text{i}}^{{3\ov2}}-n_{\text{f}}^{{3\ov2}}\ov n_{%
\text{i}}^{{3\ov2}}} & \text{case I} \\ 
&  \\ 
{1\ov2\varepsilon }{n_{\text{i}}^{2}-n_{\text{f}}^{2}\ov n_{\text{i}}^{2}} & 
\text{case II}%
\end{cases}%
,
\end{equation}%
where we have used $H^{2}\sim {8\pi G\ov3}\rho ^{V}$ and formula (\ref%
{rhovacacc}) for the time dependence of the energy density. The subscript $%
i/f$ indicates quantities at the beginning/end of the inflationary period
(in particular $n_{\text{i}}=N$ using our previous notation). It is easy to
accommodate the desired value for $N_{e}$ by choosing $n_{\text{f}}
$ not too close to $n_{\text{i}}$ (but not necessarily as low as 1 or 0) and
an appropriate value for $\varepsilon $. The specific details regarding the
end of the inflationary period (final metastable vacuum, reheating, etc.)
will, as we already have stressed, not be treated in this work. \bigskip 

\subsection{Different tunneling rates at each step of the ladder}

\label{difftunn}

\bigskip

In the case when $\widetilde{\Gamma }_{i}\neq \widetilde{\Gamma }_{j}$ in (%
\ref{systemprob}), the solution of the coupled set of differential equations
is given by 
\begin{equation}
p_{N-k}=c_{0}^{N-k}e^{-\widetilde{\Gamma }_{N-k}t}\left(
1+\sum_{i=N-k+1}^{N}c_{i}^{N-k}e^{-\left( \widetilde{\Gamma }_{N-i}-%
\widetilde{\Gamma }_{N-k}\right) t}\right) ,
\end{equation}%
where the coefficients $c_{i}^{N-k}$ are given by the recursive equations%
\begin{equation}
\left\{ \begin{aligned} c_0^{N-k+1} & = c_0^{N-k}c_{N-k+1}^{N-k}
{\left(\widetilde\Gamma_{N-k}-\widetilde\Gamma_{N-k+1}\right)\ov
\widetilde\Gamma_{N-k+1}} & \\ c_i^{N-k+1} & = c_i^{N-k}
{\left(\widetilde\Gamma_{N-k}-\widetilde\Gamma_{N-i}\right) \ov
c_{N-k+1}^{N-k}\left(\widetilde\Gamma_{N-k}-\widetilde\Gamma_{N-k+1}\right)}
& \text{for} N-k+2\leq i\leq N \\ \sum_{i=N-k+1}^N c_i^{N-k} & = -1 &
\end{aligned}\right.
\end{equation}%
As in the previous case we are interested in investigating the conditions
for which there is a slow-roll regime, and in studying the values for the
tunneling rates $\widetilde{\Gamma }_{m}$ leading to percolation and large
scale thermalization.

In general it is possible to understand how a slow-roll behaviour can arise,
by observing that every $p_{N-k}$ for $k\neq 0$ is a function peaked (how
much depending on the $\widetilde{\Gamma }_{m}$'s) at a certain time $%
t_{N-k}^{\text{max}}$. Therefore the tunnelling chain is effectively slowed
down at every step, until the fraction of volume of the new phase reaches
its maximum. It is however not possible to write a formula for the total
vacuum energy such as (\ref{vacenerfor}) unless a model is specified, so
that the precise dependence of the $\widetilde{\Gamma }_{m}$ on $m$ is known.

Nevertheless it is possible to estimate the slow-roll parameter as follows.
During slow-roll we naturally expect:  
\begin{equation}  \label{approxHslow}
H^2 \sim {8\pi G \ov 3}\rho^V
\end{equation}
and\footnote{%
We in general expect this formula because the energy  density of uncollided
walls is proportional to the energy difference  between two consecutive
vacua, while the one of the interior of  bubbles is proportional to the
energy level, which is greater than  the difference. See also the discussion
in the previous section.}  
\begin{equation}
\rho^V \sim \sum_m \epsilon_m p_m
\end{equation}
From (\ref{systemprob}) we find  
\begin{equation}
\dot \rho^V = -\sum_m \Delta\epsilon_m\widetilde\Gamma_m p_m
\end{equation}
and from (\ref{modelsenergy}) we see  
\begin{eqnarray}  \label{approxDeltaeps}
\Delta\epsilon_m & = & {\epsilon_m \ov m} \qquad \text{case I}  \notag \\
& ~ & \\
\Delta\epsilon_m & \sim & {2\epsilon_m \ov m} \qquad \text{case II}.   \notag
\end{eqnarray}
Then, using this and (\ref{approxHslow}), we find for the slow-roll
parameter  
\begin{equation}  \label{slowrolldiffratgenform}
\varepsilon = {1 \ov 2H} \times 
\begin{cases}
{\sum_m \epsilon_m\widetilde\Gamma_m p_m\,m^{-1} \ov \sum_n \epsilon_n p_n}
& \text{case I} \\ 
2{\sum_m \epsilon_m\widetilde\Gamma_m p_m\,m^{-1}\ov \sum_n \epsilon_n p_n}
& \text{case II}%
\end{cases}
\,\,.
\end{equation}
If we now define an average $\langle{\widetilde\Gamma \ov n}\rangle$ as  
\begin{equation}
\langle{\widetilde\Gamma \ov n}\rangle = {\sum_m \epsilon_m
\widetilde\Gamma_m p_m\, m^{-1} \ov \sum_n \epsilon_n p_n}.
\end{equation}
Then the slow-roll parameter is given by  
\begin{equation}
\varepsilon \equiv -\frac{\dot{H}}{H^{2}}\approx 
\begin{cases}
\frac{1}{2H} \langle{\widetilde\Gamma\ov n}\rangle & \text{case I} \\ 
\frac{1}{H\tau } \langle{\widetilde\Gamma\ov n}\rangle & \text{case II}%
\end{cases}%
\,\, .  \label{slowrollparneqrate}
\end{equation}
It is interesting (since it will be needed later on, when we will deal with
the spectral index of perturbations) to compute the time derivative of the
slow-roll parameter. From (\ref{slowrolldiffratgenform}) and using (\ref%
{approxDeltaeps}):  
\begin{equation}  \label{slowrollderiv}
\dot \varepsilon = 
\begin{cases}
{\varepsilon \ov 2} \langle{\widetilde\Gamma \ov n}\rangle+ {1 \ov 2H}\left({%
\sum_m \epsilon_m\widetilde\Gamma_m \dot p_m\,m^{-1} \ov \sum_n \epsilon_n
p_n} +{(\sum_m \epsilon_m\widetilde\Gamma_m p_m\,m^{-1})^2 \ov (\sum_n
\epsilon_n p_n)^2}\right) & \text{case I} \\ 
\varepsilon \langle{\widetilde\Gamma \ov n}\rangle+ {1 \ov H}\left({\sum_m
\epsilon_m\widetilde\Gamma_m \dot p_m\,m^{-1} \ov \sum_n \epsilon_n p_n} +2{%
(\sum_m \epsilon_m\widetilde\Gamma_m p_m\,m^{-1})^2 \ov (\sum_n \epsilon_n
p_n)^2}\right) & \text{case II}%
\end{cases}%
\end{equation}
Let us focus on the two terms in each of the brackets. The second one is
simply  
\begin{equation}
{(\sum_m \epsilon_m\widetilde\Gamma_m p_m\,m^{-1})^2 \ov (\sum_n \epsilon_n
p_n)^2}= \langle{\widetilde\Gamma \ov n}\rangle^2 \,\, .
\end{equation}
It is the easy to see that using (\ref{systemprob}), the numerator in the
first term reads  
\begin{eqnarray}
\sum_m \epsilon_m\widetilde\Gamma_m \dot p_m\,m^{-1} & = &
\sum_{m}\widetilde\Gamma_{m+1}p_{m+1} \left({\epsilon_m \ov m}%
\widetilde\Gamma_{m}- {\epsilon_{m+1} \ov m+1}\widetilde\Gamma_{m+1}\right) 
\notag \\
& = & -\sum_{m}\widetilde\Gamma_{m+1}p_{m+1} \Big(\Delta\left({%
\epsilon_{m+1} \ov m+1}\right)\widetilde\Gamma_{m+1}+ {\epsilon_{m} \ov m}%
\Delta(\widetilde\Gamma_{m+1})\Big)
\end{eqnarray}
where we have defined, for any quantity $f_m$:  
\begin{equation}
\Delta \left(f_m\right) \equiv f_m-f_{m-1} \,\,.
\end{equation}
We find that:  
\begin{equation}
\Delta\left({\epsilon_{m+1} \ov m+1}\right) = 
\begin{cases}
0 & \text{case I} \\ 
{m_f^2 \ov 2}= {\epsilon_{m+1} \ov (m+1)^2} & \text{case II}%
\end{cases}
\,\, .
\end{equation}
If we now define  
\begin{equation}
\sigma_{\langle{\widetilde\Gamma \ov n}\rangle}^2 = \left\langle\left(\text{%
{\small ${\widetilde\Gamma \ov n}$}}\right)^2\right\rangle - \left\langle%
\text{{\small ${\widetilde\Gamma \ov n}$}}\right\rangle^2 \,\,.
\end{equation}
Then, from (\ref{slowrollderiv}) and (\ref{slowrollparneqrate}), we have  
\begin{equation}  \label{derivslowrollfinal}
\dot\varepsilon \sim 
\begin{cases}
3H\varepsilon^2 - \varepsilon \langle{\widetilde\Gamma \ov n}%
\Delta\widetilde\Gamma \rangle \langle{\widetilde\Gamma \ov n}\rangle^{-1} & 
\text{case I} \\ 
2H\varepsilon^2 -\varepsilon \,\sigma_{\widetilde\Gamma\ov n}^2 \,\, \langle{%
\widetilde\Gamma \ov n}\rangle^{-1} - \varepsilon \langle{\widetilde\Gamma %
\ov n}\Delta\widetilde\Gamma \rangle \langle{\widetilde\Gamma \ov n}%
\rangle^{-1} & \text{case II} \,\,.%
\end{cases}%
\end{equation}
where all the averaging has been made using the distribution $\rho_m =
\epsilon_mp_m$. We note that in case I, the derivative of the slow-roll
parameter contains a term which will also be found in the case of equal
tunneling rates at every step (since it does not depend on the difference of
these decay rates) and a term that is depending on the differences
themselves. We call them the ``universal'' term and the ``difference'' term.

In case II, instead, we find, beyond the universal and the difference terms,
also a contribution deriving from the spread of the distribution $\rho_m$.

\bigskip


\section{Cosmological perturbations in chain inflation}

\label{secpert} 

\subsection{General considerations}

Inflationary models must be able to fulfill a series of requirements such as
providing a solution to the flatness, entropy, horizon and monopole problems
of the standard Big Bang. But at the same time, they must also meet
experimental constraints coming from measurements of the spectrum of
cosmological perturbations.

This last point represents a serious issue for chain inflation, as well as
for any inflationary model not depending on a single field. Indeed, the CMB
analysis strongly constrains the amount of entropy perturbations and the
form of the spectrum (and higher order non-gaussianities) of the adiabatic
density perturbations.

In this section we will derive a set of equations that enable us to discuss
the cosmological perturbations and compute their adiabatic spectrum.

\subsection{Scalar-type perturbations}

\bigskip

\subsubsection{The nature of perturbations}

~

Within a chain inflation model, density perturbations are generated and
evolved in a complicated way. Contrary to the usual slow-roll, chaotic
inflation model (or any model not containing first order phase transitions),
we have different kind of perturbations being produced. In a slow-roll model
the seeds of cosmological perturbations are provided only by the quantum
fluctuations that accompany the classical evolutions of the fields, metric
and fluids. Their dynamics can be described by a set of differential
equations that relate the perturbations of the different entities (metric,
fields, fluids).

Within a first order inflationary scenario, we have instead two kinds of
perturbations being generated. On the one hand the quantum fluctuations as
in the case of slow-roll, chaotic inflation, and on the other hand the
statistical fluctuations of bubble nucleation and collision. The presence of
these two seeds of cosmological perturbations makes it interesting and
important to study and analyze this aspect of the model.

It is clear that we cannot straight away say which of the two mechanism of
generation is the dominating one. Indeed, the basic quantity we use to
describe the energy density for vacuum and radiation is represented by the
fraction of universe occupied by the $m-$th vacuum 
\begin{eqnarray}
\dot{p}_{m}(t) &=&-\Gamma _{m}V^{\text{physical}}(t,t_{f})p_{m}(t)+\Gamma
_{m+1}V^{\text{physical}}(t,t_{f})p_{m+1}(t)  \notag  \label{equatprob} \\
&\equiv &-\widetilde{\Gamma }_{m}(t)p_{m}(t)+\widetilde{\Gamma }%
_{m+1}(t)p_{m+1}(t),
\end{eqnarray}%
The quantity $\widetilde{\Gamma }_{m}(t)$ represents the average number of
bubbles of the $m$-th vacuum generated at a time $t$, irrespective of the
bubbles generated at previous times. The actual number of such bubbles is
therefore distributed according to a Poisson distribution with mean value $%
\widetilde{\Gamma }_{m}(t)$. Therefore, a first kind of fluctuation in $%
p_{m}(t)$ is due to the statistical fluctuations in this quantity, being
equal to $\sqrt{\widetilde{\Gamma }_{m}(t)}$. In general, solving for $%
p_{m}, $ we can determine the perturbation in the energy density due to
these statistical fluctuations.

But this is not the only seed of fluctuations in $p_{m}(t)$. Also, the
volume $V^{\text{physical}}(t,t^{\prime })$ depends on the metric which it
is measured by. And therefore the quantum fluctuations in the metric field
generate and are sourced by perturbations in the energy density of the
vacuum and of the radiation as well, due to the coupling between these two.

Our goal will therefore be to obtain a set of equations for the
perturbations (relating those of the metric and the various energy and
pressure densities), to solve them and to obtain the spectrum of the
fluctuations.

We will use the \emph{longitudinal gauge}, where the most general metric
with scalar perturbations can be written 
\begin{equation}
ds^{2}=-(1+2\phi )d\,t^{2}+2a\partial _{i}Bd\,td\,x^{i}+((1-2\psi )\delta
_{ij}+D_{ij}E)d\,x^{i}d\,x^{j}),  \label{metripertscal}
\end{equation}%
with $B=E=0$. Note that here $D_{ij}\equiv \partial _{i}\partial _{j}-{1\ov3}%
\delta _{ij}\nabla ^{2}$.

Let us now pause for a moment to discuss the general approach to
cosmological perturbations. The distinction that is made in cosmology
between perturbed and unperturbed universes is a fictitious one. The actual
universe has a metric of the form (\ref{metripertscal})\footnote{%
We consider only scalar perturbations for the moment.}, and we compare it
with a fictitious universe where the metric is a perfect Robertson-Walker
one. In our case, $p_{m}$ has a well defined geometrical meaning independent
of the metric used to compute $\widetilde{\Gamma }_{m}(t),$ and has to obey
the equation 
\begin{equation}
\dot{p}_{m}=-\widetilde{\Gamma }_{m}p_{m}+\widetilde{\Gamma }_{m+1}p_{m+1},
\label{widep}
\end{equation}%
both in the perturbed and unperturbed case. It is therefore possible to
determine the perturbations in $p_{m}$ just by writing $\widetilde{\Gamma }$
in terms of the perturbed fields and find the quantum part (for example in $%
V^{\text{physical}}(t,t^{\prime })$) and adding the relevant statistical
perturbation. Similarly, we find the form of $\mathcal{F}^{m+1,m}$ to be
determined by the geometry also in the presence of perturbations.

\subsubsection{Notation and basic quantities}

\bigskip

As before it is convenient to formally distinguish every phase with an index 
$m,$ and associate to each one a vacuum and radiation stress-energy tensor,
distinguished by $\ell =V,r$ as in formulas (\ref{backlversmpart}). We will
also distinguish the quantities in the perturbed (real) universe form those
in the \textquotedblleft background\textquotedblright\ one by writing the
first one with a bar, the second one without. Hence we write 
\begin{equation}
\begin{cases}
\bar{A} & \text{is the perturbed value} \\ 
A & \text{is the background}%
\end{cases}%
,
\end{equation}%
with%
\begin{equation*}
\bar{A}=A+\delta A+O((\delta A)^{2}).
\end{equation*}%
In this formalism the fundamental equations governing the dynamics of the
various generalized fluids are 
\begin{equation}
\nabla ^{\mu }\bar{T}_{\mu \nu }^{\ell ,m}=\bar{Q}_{\nu }^{\ell ,m},
\label{Tmunueq}
\end{equation}%
with, in order for the total stress-energy tensor to be conserved, 
\begin{equation}
\sum_{\ell ,m}\bar{Q}_{\nu }^{\ell ,m}=0\quad \Rightarrow \quad \sum_{\ell
,m}Q_{\nu }^{\ell ,m}=\sum_{\ell ,m}\delta Q_{\nu }^{\ell ,m}=0.
\label{conservtot}
\end{equation}

Let us focus, since it will be specially important in the following, on the
vacuum sector. First of all, the energy momentum tensor is given by 
\begin{equation}
\bar{T}_{\mu \nu }^{V}=\sum_{m}(\bar{\rho}^{V,m}+\bar{P}^{V,m})\bar{u}_{\mu
}^{V}\bar{u}_{\nu }^{V}+\bar{P}^{V,m}\bar{g}_{\mu \nu },
\label{enmomtenstot}
\end{equation}%
where $\bar{u}^{V}$ is the proper velocity for the vacuum, and therefore we
have 
\begin{equation}
\bar{u}^{V,\alpha }\bar{T}_{\alpha \beta }^{V,m}=\bar{\rho}^{V,m}\bar{u}%
_{\beta }^{V}.  \label{entenseig}
\end{equation}%
Second, recall that the formula for the vacuum energy at every phase is
given by 
\begin{equation}
\rho ^{V,m}=\rho ^{\mathcal{V},m}+\rho ^{W,m},
\end{equation}%
where $\rho ^{\mathcal{V},m}$ is the energy in the interior of the bubbles
of the $m$-th phase and $\rho ^{W,m}$ is the energy stored within the walls
between phase $m+1$ and phase $m$. We also know that the dynamical evolution
of the energy density is 
\begin{equation}
\dot{\bar{\rho}}^{V,m}=-\bar{\widetilde{\Gamma }}_{m}\bar{\rho}^{\mathcal{V}%
,m}+\bar{\widetilde{\Gamma }}_{m+1}\bar{\rho}^{\mathcal{V},m+1}-\bar{%
\widetilde{\gamma }}_{m}\bar{\rho}^{W,m},
\end{equation}%
where $\bar{\tilde{\gamma}}_{m}$ represents the conversion of the energy of
the walls into radiation\footnote{%
At the microscopical level this would be a series of complicated processes
of collisions of walls and decays, but we will instead use a well-defined
coarse-grained description.}.

By comparing with (\ref{Tmunueq}), it is then straightforward to find that
the covariant form of $\bar{Q}_{\nu }^{V,m}$ is: 
\begin{eqnarray}
\bar{Q}_{\nu }^{V,m} &=&-\bar{\widetilde{\Gamma }}_{m}\,\bar{u}^{V\alpha }\,%
\bar{T}_{\alpha \nu }^{V,m}+\bar{\widetilde{\Gamma }}_{m+1}\,\bar{u}%
^{V\alpha }\,\bar{T}_{\alpha \nu }^{V,m+1}-\bar{\widetilde{\gamma }}_{m}\,%
\bar{u}^{V\alpha }\,\bar{T}_{\alpha \nu }^{W,m}  \notag  \label{QVmdef} \\
& \equiv &\bar{Q}^{V,m}\bar{u}_{\nu }^{V}
\end{eqnarray}%
where in the last line we have used (\ref{entenseig}). Note also that $%
u^{V\alpha }$ is the same for all components with different $m,$ since it is
just the proper velocity of the total vacuum, whose energy momentum tensor
is (\ref{enmomtenstot}).

The perturbations in the sources $Q^{\ell ,m}$ can be written as \cite%
{Malik:2004tf, KodSas, Hwang:2001qk, Hwang:2001fb} 
\begin{eqnarray}
\delta Q_{0}^{\ell ,m} &=&-Q^{\ell ,m}\phi -\delta Q^{V,m} \\
\delta Q_{i}^{\ell ,m} &=&\partial _{i}\left( Q^{\ell ,m}v+f^{\ell
,m}\right) ,  \label{transQpert}
\end{eqnarray}%
where $v$ is the velocity potential, obtained by expanding the spatial
components of the average proper velocity $u^{i}$ in harmonic functions as
in \cite{KodSas, Hwang:2001qk}. In our case, for $\ell =V,r,$ we find 
\begin{equation}
v=\sum_{\ell ,m}{\rho ^{\ell ,m}+P^{\ell ,m}\ov\rho +P}v^{\ell
,m}=\sum_{m}v^{r,m}=v^{r}.  \label{averagev}
\end{equation}%
For ease of notation, we will also use 
\begin{equation}
\delta q\equiv (p+P)v,
\end{equation}%
where  
\begin{eqnarray}
\delta q &\equiv &\sum_{\ell }\delta q^{\ell }  \label{totverspart} \\
\delta q^{\ell ,m} &\equiv &(\rho ^{\ell ,m}+P^{\ell ,m})v^{\ell ,m}.
\label{partqversvel}
\end{eqnarray}%
By expanding (\ref{QVmdef}) in background plus perturbations, writing the
proper velocity in terms of the velocity potential, comparing with (\ref%
{transQpert}) and using (\ref{averagev}) we find for the vacuum 
\begin{equation}
Q^{V,m}v^{V}=Q^{V,m}v+f^{V,m}=Q^{V,m}v^{r}+f^{V,m}.
\end{equation}

In order to study entropy (isocurvature) perturbations, it is convenient to
define the gauge invariant relative entropy perturbation between fluids $%
\ell $ and $\ell ^{\prime }$, 
\begin{equation}
S_{\ell ,\ell ^{\prime }}=-3\left( \mathcal{R}_{\ell }-\mathcal{R}_{\ell
^{\prime }}\right) .
\end{equation}%
In the usual definition, for hydrodynamical matter, $S_{\ell ,\ell ^{\prime
}}$ is defined in terms of the curvature perturbation on uniform density
hypersurfaces. Since we are now studying a system where many scalar fields
play a role, the correct definition (see \cite{postLiMuSa}) is in terms of
the \emph{comoving curvature perturbation}\,\footnote{%
In  \cite{postLiMuSa} this is indicated as $\zeta$, which is  unfortunately 
also used  for the curvature perturbation on uniform energy density
hypersurfaces.} 
\begin{equation}
\mathcal{R}_{\ell }=\phi -Hv_{\ell }.
\end{equation}%
The total density perturbation is given by (\cite{Malik:2004tf}): 
\begin{equation}
\mathcal{R}=\phi -H\,v.
\end{equation}%
By using these definitions and (\ref{averagev}), we find: 
\begin{equation}
f^{V}=\sum_{m}f^{V,m}=Q^{V}(v^{V}-v^{r})={Q^{V}\ov3H}S_{V,r}  \label{fventr}
\end{equation}

\subsubsection{Equations of motion and spectrum of perturbations}

~

The most general set of equation for a set of generalized fluids (including
scalar fields) for the case of pure Einsten-Hilbert theory has been written
in \cite{Hwang:2001fb} (for generalized theories of gravity, the enlarged
set of equations is in \cite{Hwang:2001qk}). In particular we will use (see
also \cite{Malik:2004tf}): 
\begin{align}
& \psi =\phi \\
& -{k^{2}\ov a^{2}}\phi -3H(H\phi +\dot{\phi})=4\pi G\delta \rho
\label{Irho} \\
& (H\phi +\dot{\phi})=-4\pi G(\rho +p)v=-4\pi G\delta q  \label{Iv} \\
& \dot{\delta \rho }^{\ell ,m}+3H(\delta \rho ^{\ell ,m}+\delta P^{\ell
,m})=3(\rho ^{\ell ,m}+P^{\ell ,m})\Big(\dot{\phi}+{k\ov a}v^{\ell, m}\Big)%
+Q^{\ell ,m}\phi+\delta Q^{\ell ,m}  \label{rhopart} \\
& \dot{\delta q}^{\ell ,m}+3H\delta q^{\ell ,m}+(\rho ^{\ell ,m}+P^{\ell
,m})\phi +\delta P^{\ell ,m}=Q^{\ell ,m}{\delta q\ov\rho +p}+f^{\ell ,m}
\label{qpart} \\
& \dot{\delta q}+3H\delta q+(\rho +P)\phi +\delta P=0  \label{qtot}
\end{align}

Note that $\delta \rho ^{\ell ,m},\,\delta p^{\ell ,m}$, etc, include the
contributions from quantum as well as statistical fluctuations. In order to
re-write this system of equations in a useful way, we need to establish the
relation between the total pressure and energy densities. We already know
that 
\begin{equation}
\delta P^{r}={1\ov3}\delta \rho ^{r},
\end{equation}%
but we also need to find the relation between the corresponding quantities
for the vacuum energy. This can be done by inspecting the relevant equation
of motion. From (\ref{Tmunueq}), we have: 
\begin{equation}
\dot{\rho}^{V,m}=Q^{V,m}.
\end{equation}%
Since, according to our discussion above, this relation is valid both for
the unperturbed and for the perturbed universes, we immediately obtain 
\begin{equation}
\dot{\delta}\rho ^{V,m}=\delta Q^{V,m}.
\end{equation}%
By comparing this result with the general equation (\ref{rhopart}) for $\ell
=V$, we find 
\begin{equation}
\delta P^{V,m}=-\delta \rho ^{V,m}+{Q^{V,m}\ov3H}\phi ,
\end{equation}%
and therefore 
\begin{eqnarray}
\delta P^{V} &=&\sum_{m}\delta P^{V,m}=-\delta \rho ^{V}+\sum_{m}{Q^{V,m}\ov%
3H}\phi  \notag  \label{dpV1} \\
&=&-\delta \rho +\delta \rho ^{r}+\sum_{m}{Q^{V,m}\ov3H}\phi .
\end{eqnarray}

If we compare this with the result we would obtain in a standard single
field slow roll or chaotic scenario, we see the importance of the presence
of perturbations of the radiation at every step of the tunnelling chain. In
fact, $\delta \rho ^{r}$ plays here a role similar to the perturbation of
the scalar field kinetic energy in standard slow roll. It is therefore
important to keep track of it, as it will contribute substantially to the
dynamical equation of the perturbations.

On the other hand, from the definition of the momentum perturbation in (\ref%
{partqversvel}), we see that 
\begin{equation}
\delta q^{V,m}=0.
\end{equation}%
Therefore, using (\ref{qpart}) for $\ell =V$, we find 
\begin{equation}
\delta P^{V,m}={Q^{V,m}\ov3H (\rho+P)}\left( \delta \rho +{k^{2}\ov4\pi
Ga^{2}}\phi \right) +f^{V,m},
\end{equation}%
and so, summing over all $m$'s: 
\begin{equation}
\delta P^{V}={Q^{V}\ov3H (\rho+P)} \left(\delta \rho +{k^{2}\ov4\pi Ga^{2}}%
\phi\right) +f^{V}.  \label{dpV2}
\end{equation}%
Then, from (\ref{dpV1}, \ref{dpV2}, \ref{fventr}) and neglecting a
sub-leading term $\propto \dot{\rho}^r$, we get 
\begin{equation}
\delta \rho ^{r}=-\sum_{m}{Q^{V,m}\ov3H}\phi -{k^{2}\ov4\pi Ga^{2}}\phi
+\sum_{m}{Q^{V,m}\ov H}S_{V,r},  \label{deltarhorad}
\end{equation}%
where we have used: 
\begin{equation}
{4\pi G\ov 3H}\sum_{m}Q^{V,m}=4\pi G{\dot{\rho}^{V}\ov 3H}={4\pi G\ov 3H}{d%
\ov dt}({3(2-\epsilon )\ov16\pi G}H^{2}) \sim \dot{H} .  \label{QVslowroll}
\end{equation}%
As a consequence 
\begin{eqnarray}  \label{totdeltaPfin}
\delta P &=&\delta P^{V}+\delta P^{r}  \notag  \label{deltaPtot} \\
&=&-\delta \rho -{4\ov3}{k^{2}\ov4\pi Ga^{2}}\phi -{1\ov3}\sum_{m}{Q^{V,m}\ov%
3H}\phi +{4\ov3}\sum_{m}{Q^{V,m}\ov3H}S_{V,r}.
\end{eqnarray}%
Using (\ref{QVslowroll}) together with (\ref{totdeltaPfin}), we finally get
the system of equations (from (\ref{Iv}) and (\ref{qtot}))  
\begin{equation}
\left\{ \begin{aligned} (a\phi)^{^{\mathbf{.}}} & = -4\pi G\, a\, \delta q
\\ & \\ {\dot{\delta q}\ov\rho +p} & = - \left(1-{1\ov3}{k^{2}\ov 4\pi G
a^2(\rho+P)}-{1\ov3}{\dot H \ov 4\pi G (\rho+P)}\right)\phi & \\ & \quad
-{4\ov3}{\dot H \ov 4\pi G (\rho+P)}S_{V,r} & \end{aligned}\right.
\end{equation}

It is convenient to introduce the new variables $\mathcal{R},\xi $ defined
by 
\begin{eqnarray}
a\phi &=&4\pi GH\xi \\
&&  \notag \\
{\delta q\ov\rho +p} &=&-{\mathcal{R}\ov H}+{4\pi G\ov a}\xi ,
\end{eqnarray}%
where we easily recognize in $\mathcal{R}$ the \emph{comoving curvature
perturbation}\footnote{%
In \cite{kinflGarrMuk} this is indicated as $\zeta$.}. In terms of these new
variables, the system becomes 
\begin{equation}
\left\{ \begin{aligned} \dot\xi & = {a(\rho+p) \ov H^2}\mathcal{R} \\
\dot{\mathcal{R}} & = {1 \ov 3} {H^2 \ov a^3 (\rho+p)}\big(-k^2 +a^2
H^2\varepsilon\big)\xi -{4 \ov 3} H S_{V, r} \end{aligned}\right. .
\label{systpert}
\end{equation}

For the moment we concentrate on the adiabatic perturbations, neglecting the
term proportional to $S_{V,r}$. Non-adiabatic isocurvature perturbations (as
well as non-gaussianities\,\footnote{%
See for example \cite{ChWaXu}.}) are expected to be present in chain
inflation models because of the presence of various fields that can follow
trajectories in field space different from the inflaton one, but this
depends strongly on the specific features of the various models, which our
very general picture does not accommodate.

General arguments justifying the negligibility of the isocurvature
perturbations are often found in the literature, but they do not apply in
our case, or in any other case with coexisting fields and conventional
hydrodynamical fluids that interact. Lacking a general argument regarding
the amount of the isocurvature perturbations, we would need to rely on
specific analysis of the detailed models. This we do not attempt to do, and
our conclusions in what follows are valid only in those cases were the
detailed model does indeed contain a negligible contribution of isocurvature
perturbations.

As a word of comment, note that the difference between our final result and
the usual one (for non-interacting fields and fluids) is, apart from the
entropy term, proportional to ${Q^{V,m}\ov3H}\sim \varepsilon $ (see (\ref%
{deltaPtot})). It is therefore, as one could expect, due to the interactions.

In order to normalize the amplitude of quantum fluctuations, we need the
action governing their dynamics. Our setup, being so general, lacks this
information. Nevertheless, following \cite{kinflGarrMuk}, we can infer the
action from the equations of motion (\ref{systpert}) themselves, up to a
time independent factor. Following the same steps as in \cite{kinflGarrMuk},
we introduce the canonical quantization variable $\varsigma =z\mathcal{R}$
with 
\begin{equation}
z\equiv {a(\rho +p)^{1\ov2}\ov {1 \ov 3}H}\left( {\hat{O}\ov(-k^{2}+\mathcal{%
H}^{2}\varepsilon )}\right) ^{1\ov2}
\end{equation}%
where we have used the conformal time defined as $\eta =\int a^{-1}dt$. The
time independent operator $\hat{O}$ can be determined by comparing our
result with well-known ones (for example with the pure radiation case). We
find then that $\hat{O}=-k^{2}$. The action for the variable $\varsigma $ is
in our case found to be (up to a total derivative term) 
\begin{equation}
S={1\ov2}\int \big[\varsigma ^{^{\prime }2}+{1\ov3}\varsigma (\Delta +%
\mathcal{H}^{2}\varepsilon )\varsigma +{z^{^{\prime \prime }}\ov z}\varsigma
^{2}\big]d\!\eta d^{3}\!x,
\end{equation}%
and the equation of motion is 
\begin{equation}
\varsigma ^{^{\prime \prime }}+\left( {1\ov3}k^{2}-{1\ov3}\mathcal{H}%
^{2}\varepsilon -{z^{^{\prime \prime }}\ov z}\right) \varsigma = 0 ,
\end{equation}%
with the obvious long wavelength solution $\varsigma \sim z\rightarrow 
\mathcal{R}\sim const$.

Since ${z^{^{\prime \prime }}\ov z}\sim {a^{^{\prime \prime }}\ov a}\sim
2(Ha)^{2}$ we find\footnote{%
We use the reduced Plank mass $M_{\text{pl}}=\left( 8\pi G\right) ^{-1}$.}
following \cite{kinflGarrMuk} 
\begin{equation}
P_{k}^{\mathcal{R}}={H^{2}\ov8\pi ^{2}M_{Plank}^{2}{1\ov \sqrt{3}}%
\varepsilon ,}
\end{equation}%
for the spectrum of perturbations, where quantities have to be evaluated at
a time given by $aH={k\ov \sqrt{3}}$.

\bigskip


\section{A potential for chain inflation from flux compactification}

\label{secmodel} 

In order for chain inflation to be a serious alternative to chaotic
inflation, one needs a framework where the required potential arises in a
natural way. As we will see such a framework is provided by flux
compactified string theory.

\subsection{Series of minima from type IIB flux compactifications}

\bigskip

A particularly interesting compactification of string theory is type IIB
with fluxes. It serves as a useful playground for constructing models of the
early universe, and exhibits in a rather explicit way the problems and
promises of the string landscape.

The potential for a flux compactified type IIB string is given by%
\begin{equation}
V\left( z,\tau \right) =e^{K}g^{i\bar{\imath}}D_{i}WD_{\bar{\imath}}\bar{W}%
+e^{K}g^{\tau \bar{\tau}}D_{\tau }WD_{\bar{\tau}}\bar{W}-3e^{K}\left\vert
W\right\vert ^{2},
\end{equation}%
where $W$ is the superpotential and $K$ is the K\"{a}hler potential. We will
focus on the dependence on the complex structure moduli and are therefore
mostly interested in the structure of the superpotential. It is obtained
from the fluxes as%
\begin{equation}
W=F\cdot \Pi -\tau H\cdot \Pi ,
\end{equation}%
where the period vectors are defined to be $\Pi =(\int_{A_{1}}\Omega
,\int_{B_{1}}\Omega ,...)^{T},$ with $\Omega $ as the holomorphic three form.

As shown in \cite{DanJohLar, ChDaJoLaVo} it is rather easy to find fluxes
where the potential has a minimum somewhere in moduli space. Of particular
interest are minima where one of the components of the flux vector, say $%
F_{2}$, is much smaller than the corresponding conjugated component, i.e. $%
F_{1}$. We can then make use of conifold monodromies to generate new minima.
Specifically, we need to move in moduli space around the conifold point
where the cycle $A_{1}$ vanishes. When we do that we go through a cut and
end up on a new sheet. Furthermore, the cycles transform as $%
B_{1}\rightarrow B_{1}+nA_{1}$, where $n$ is the number of times we go
around the conifold. The potential after the transformation can equivalently
be evaluated using the old periods but with%
\begin{align}
F_{(3)}& \rightarrow F_{(3)}+n\left( F_{2},0,0,...,0\right) \\
H_{(3)}& \rightarrow H_{(3)}+n\left( H_{2},0,0,...,0\right) .
\end{align}%
If $F_{2}\ll F_{1}$we can be reasonably sure that we will find a new minimum
near the old one on the next sheet. In this way we can generate an infinite
series of minima if the $F_{1}\rightarrow \infty $ limit of $F_{(3)}$
corresponds to a potential with a minimum. The minima will be positioned
along a spiral staircase turning around the conifold point.

In \cite{DanJohLar} it was discussed in detail how to obtain such series. In 
\cite{ChDaJoLaVo} it was suggested that one could make use of geometric
transitions between different Calabi-Yau manifolds to help generate the
series.

\bigskip

\subsection{A potential for chain inflation}

\bigskip

Instead of working with explicit flux compactifications for specific
Calabi-Yau manifolds, we will capture the essential features of the
potential with a simple toy model. The main results will depend on just a
limited number of parameters that in turn will depend on the complex
geometry of the actual compactification.

The value of the potential at the minima can effectively be described using
the results of \cite{DanJohLar} and general properties of the potential used
in type IIB flux compactifications. There are two cases that are of
particular interest. In \cite{DanJohLar} it was found that when the
axiodilaton modulus is fixed by the fluxes themselves the potential at the
minima has a form such that 
\begin{equation}
V_{n}=m_{f}^{4}n=m^{3}\varphi _{n}\qquad \text{Case I}.
\end{equation}%
If the axiodilaton is fixed by other mechanisms we instead expect 
\begin{equation}
V_{n}=\frac{1}{2}m_{f}^{4}n^{2}=\frac{1}{2}m^{2}\varphi _{n}^{2}\qquad \text{%
Case II}.
\end{equation}%
$n$ is in both cases given by the flux $F_{1},$ and we have assumed that we
are in a limit where the flux is large but the the minima still remain. In
general the potential $V$ is the sum of an underlying $m^{3}\varphi $ (case
I) or $\frac{1}{2}m^{2}\varphi ^{2}$ (case II) and superimposed barriers.
What we have found in this way, is a realization of the chain inflation
potential studied in the previous section.

It immediately follows that the energy difference between two adjacent
minima is given by 
\begin{equation}
\begin{cases}
\Delta V=m_{f}^{4} & \text{case I} \\ 
\Delta V=m_{f}^{4}n & \text{case II}%
\end{cases}%
\end{equation}%
We can also calculate the field values at the minima with the result 
\begin{equation}  \label{fieldatmin}
\varphi =%
\begin{cases}
\frac{m_{f}^{4}}{m^{3}}n & \text{case I} \\ 
\frac{m_{f}^{2}}{m}n & \text{case II},%
\end{cases}%
\end{equation}%
and the distance in field space between two adjacent minima is given by 
\begin{equation}
\Delta \varphi =%
\begin{cases}
\frac{m_{f}^{4}}{m^{3}} & \text{case I} \\ 
\frac{m_{f}^{2}}{m} & \text{case II}.%
\end{cases}%
\end{equation}%
As we see, there are two energy scales involved, $m$ and $m_{f}$, that
ultimately depend on the details of the compactification. $m_{f}$ is given
by parameters involving coupling strengths and the size of the extra
dimensions, while $m$ crucially depends on details of the complex geometry
determining the distance in field space of the minima from a conifold point.
In the following we will treat $m_{f}$ and $m$ as free parameters.

Let us now turn to the tunneling. In our model the Coleman-Callan decay is
dominant, hence we have that the decay rate per unit time and physical
volume is\footnote{%
Recall that the decay rate per unit time is given by  $\widetilde{\Gamma}%
_n=\Gamma _{n}\times \text{physical volume}$.} 
\begin{equation}  \label{colcaldec}
\Gamma _{n}\sim e^{-B}= e^{-\frac{27\pi ^{2}}{2}\frac{S^{4}}{\left( \Delta
V\right) ^{3}}}.
\end{equation}%
\begin{equation}
S=\int d\varphi \sqrt{V}
\end{equation}%
is taken over the path that minimizes the integral. In our discussion of
chain inflation we have assumed a potential depending on a single (real)
scalar field. The potentials generated from flux compactifications are far
more complicated. Even if we restrict our attention to just one complex
structure moduli, the problem is essentially two dimensional with a series
of minima winding around the conifold point. We will not attempt to perform
a complete analysis of the problem, instead we will focus on a simple order
of magnitude estimate of the ingoing quantities.

We assume that the least action will essentially be given by a path
encircling the conifold point with a length given by the circumference.
Since, for large $n$, the potential $V$ scales like $n$ or $n^{2}$, it
follows that the barriers between the two minima have a height of the order 
\begin{equation}
\lambda \approx 
\begin{cases}
m_{f}^{4}n & \text{case I} \\ 
m_{f}^{4}n^{2} & \text{case II.}%
\end{cases}%
\end{equation}%
As a consequence we find 
\begin{equation}
S\sim \Delta \varphi \lambda ^{1/2}\sim 
\begin{cases}
\frac{m_{f}^{6}}{m^{3}}\sqrt{n} & \text{case I} \\ 
\frac{m_{f}^{4}}{m}\,\,n & \text{case II},%
\end{cases}%
\end{equation}%
and the exponent becomes 
\begin{equation}
B=%
\begin{cases}
\frac{27\pi ^{2}}{2}\left( \frac{m_{f}}{m}\right) ^{12}n^{2} & \text{case I}
\\ 
\frac{27\pi ^{2}}{2}\left( \frac{m_{f}}{m}\right) ^{4}n & \text{case II}.%
\end{cases}
\label{exponent}
\end{equation}%
In order for the tunneling not to be too suppressed, and $\widetilde{\Gamma}%
_n=\Gamma _{n}\times \text{physical volume}$ to be larger than one, we need
that $B$ is of order one.

\bigskip


\section{Matching cosmological data}

\label{matchdata} 

We have found that the primordial spectrum is of the form\footnote{%
This is valid for models where the amount of isocurvature perturbations is
negligible compared to that of the adiabatic ones. As we have pointed out,
there is no general argument to establish when this is the case.} 
\begin{equation}
\Delta _{\mathcal{R}}^{2}=\frac{H^{2}}{8\pi ^{2}M_{Pl}^{2}\frac{1}{\sqrt{3}}%
\varepsilon }.
\end{equation}%
This leads to the spectral index 
\begin{eqnarray}
n_{s} &=&1+\frac{1}{H}\frac{(\Delta _{\mathcal{R}}^{2})^{^{.}}}{\Delta _{%
\mathcal{R}}^{2}}=1-2\varepsilon -{\dot{\varepsilon}\ov H\varepsilon }= 
\notag  \label{genspectr} \\
&=&%
\begin{cases}
1-5\varepsilon +{1\ov H}\langle {\widetilde{\Gamma }\ov n}\Delta \widetilde{%
\Gamma }\rangle \langle {\widetilde{\Gamma }\ov n}\rangle ^{-1} & \text{case
I} \\ 
1-4\epsilon +{1\ov H}\,\sigma _{\widetilde{\Gamma }\ov n}^{2}\,\,\langle {%
\widetilde{\Gamma }\ov n}\rangle ^{-1}+{1\ov H}\langle {\widetilde{\Gamma }%
\ov n}\Delta \widetilde{\Gamma }\rangle \langle {\widetilde{\Gamma }\ov n}%
\rangle ^{-1} & \text{case II},%
\end{cases}%
\end{eqnarray}%
where we have used formula (\ref{derivslowrollfinal}).

Note that in the case of equal and constant tunneling rate for all tunneling
events, we get the result 
\begin{equation}
n_{s}= 
\begin{cases}
1-5\varepsilon & \text{case I} \\ 
1-4\epsilon & \text{case II}.%
\end{cases}%
\end{equation}%
For case II this is identically the same result as for the case of slow-roll
with a quadratic potential. The reason for this agreement is that the
quadratic potential in slow-roll models is special in the sense that it
gives rise to dynamics where the scalar field rolls with a constant speed.
This corresponds perfectly to our case II with constant tunneling time $\tau 
$. For other slow-roll models with different potentials, such as the linear
potential of case I, the results based on equal tunneling time will differ
from ordinary slow-roll based on Hubble friction.

In general, in order to satisfy the requirements on the spectral index, we
expect the contributions in (\ref{genspectr}) due to $\sigma_{\widetilde%
\Gamma / n}^2$ and to the difference in tunneling rates to be of order $%
\varepsilon$ at most. As a confirmation, it is possible to see that using
formula (\ref{colcaldec}), we would obtain a contribution $B\varepsilon $ to
the spectral index from the dependence on $n$ of the exponent. There is also
a further dependence on $n$ in the prefactor of the potential that is more
tricky to derive. We will not attempt to do this. We just note that given
the slow-roll, the corrections to the spectral index have to be fractions of 
$\varepsilon$. Let us use this for a rough estimate of the ingoing
parameters.

Matching the observations requires\footnote{%
We take the data from the 5-years WMAP survey, \cite{WMAP5}.} 
\begin{equation}
\Delta _{\mathcal{R}}^{2}=\eta ,
\end{equation}%
where $\eta \sim 2.5\cdot 10^{-9}$. The Hubble constant is then given by  
\begin{equation}
H= \sqrt{8 \pi^2 \,\eta \, {\varepsilon \ov \sqrt{3}}}\,M_{Pl}
\end{equation}
which is way below the Planck mass. Writing the Hubble constant in terms of
the energy density, leads to 
\begin{equation}
m_{f}= 
\begin{cases}
(8\sqrt{3} \pi^2 \,\eta \,\varepsilon)^{1 \ov 4} \, n^{-{1 \ov 4}} \,M_{Pl}
& \text{case I} \\ 
(8\sqrt{3} \pi^2 \,\eta \,\varepsilon )^{1 \ov 4} \,2^{1 \ov 4}\,n^{-{1 \ov 2%
}} \,M_{Pl} & \text{case II}%
\end{cases}%
\end{equation}%
and therefore $m_{f}$ is also below the Planck scale.

To estimate the parameter $m$ we must use the expression for the tunneling
amplitude, (\ref{exponent}). We find that 
\begin{equation}
m= 
\begin{cases}
({27\pi ^{2} \ov 2B})^{{1\ov 12}}n^{\frac{1}{6}}m_{f} & \text{case I} \\ 
({27\pi ^{2} \ov 2B})^{{1\ov 4}}n^{\frac{1}{4}}m_{f} & \text{case II}%
\end{cases}%
\end{equation}

Since measurements indicate $n_{s}\sim 0.97$, we must have $\varepsilon \sim
10^{-2}$. With $\varepsilon =10^{-2}$, the normalization of the spectrum
requires $\frac{H}{M_{pl}}\sim 2.6\cdot 10^{-5}$. With $n\sim 10^{4}$ we
find 
\begin{equation}
m_{f}\sim 
\begin{cases}
7.6\cdot 10^{-4}M_{pl} & \text{case I} \\ 
9\cdot 10^{-5}M_{pl} & \text{case II}%
\end{cases}%
\end{equation}%
and with $B\sim 1$ we find  
\begin{equation}
m \sim 
\begin{cases}
5.3\cdot 10^{-3}M_{pl} & \text{case I} \\ 
3.1\cdot 10^{-3}M_{pl} & \text{case II}%
\end{cases}%
\end{equation}%
These values for $m$ and $m_{f}$ are quite reasonable and can easily be
lowered by increasing $n$ further\footnote{%
It is possible then to check that with these values, in order to provide the
correct number of e-foldings and satisfy the requirements of percolation,
large scale thermalization and contraction of volume at every phase,  the
inflationary period ends before the field tunnels to the very last minima
(with vacuum numbers 1 or 0).}.

It is also of interest to note that the field values at the highest minima
are quite small, from (\ref{fieldatmin}) 
\begin{equation}
\varphi _{n}=\sim 
\begin{cases}
2.2\cdot 10^{-2}M_{pl} & \text{case I} \\ 
2.7\cdot 10^{-2}M_{pl} & \text{case II}%
\end{cases}%
.
\end{equation}%
Contrary to the case of slow-roll in chaotic inflation, we see that we need
not move over large regions in field space to achieve inflation. Comparing
with chaotic inflation one finds that the average motion of the field is in
fact much slower than would be the case in a potential of the form $\frac{1}{%
2}m^{2}\varphi ^{1, 2}$ for the above field values. The reason for the
slow-roll is the presence of the barriers that make the field temporarily
stuck before it tunnels. The barriers take the role of the Hubble friction
term, which otherwise is too weak for slow-roll. To be precise, the average
motion of the field $\varphi $ in chain inflation corresponds to $\dot{%
\varphi}^{2}\sim \varepsilon ^{2},$ rather than a single $\varepsilon $ as
in the slow roll regime of chaotic inflation. In chain inflation we instead
have a contribution from radiation, $\rho _{r}\sim \varepsilon$, that takes
over the role of the kinetic term. We should also add that it is
straightforward to verify that we obtain the correct number of e-foldings
with the parameters we are using.

It is interesting to note that the different speed of sound appearing in the
expression for $\Delta _{\mathcal{\mathcal{R}}}^{2}$ effectively increases
the strength of the scalar modes compared with the tensor modes with a
factor three. On the other hand, the chain inflation scenario sources
additional gravitational waves through the collisions of the bubble walls,
beside those due to quantum fluctuations in the metric field (which are the
only ones present in the usual slow-roll chaotic scenario). These effects
constitute an observational signature of our model. However, to really
verify them, one must analyse in details the physics of bubble collisions
and derive a precise relation between the spectral index and $\varepsilon $.
We expect this to be extremely model dependent and defer such an analysis to
future work.


\section{Conclusion}


In this paper we have found that chain inflation can be a viable model of
inflation\footnote{%
In order to establish this in a complete and rigorous way, it is still
necessary to study the reheating stage and the actual contribution and
presence of isocurvature perturbations, which we leave for future research,
when more detailed models will be available.}. We have established that a
series of tunneling events can give rise to a slowly changing vacuum energy
much in the same way as Hubble friction gives rise to slow-roll in chaotic
inflation. We have calculated the spectrum of adiabatic perturbations and
found it to be very similar to the one of chaotic inflation up to a
difference in the effective speed of sound. This is due to the fact that
radiation plays a similar role in chain inflation as the kinetic term for
the inflaton does in chaotic inflation.

We have also found that general properties of the type IIB string landscape
can be used to generate chain inflation. The key is special properties of
flux compactifications that naturally give rise to series of minima useful
for chain inflation. The relevant potentials are not flat or fine-tuned to
be flat. Instead, one carefully needs to choose the value of other
parameters that determine the amplitude for tunneling between different
vacua. Nevertheless it is interesting to see that one need not have
particularly extreme values to achieve inflation.

In order to estimate the likelihood of finding an appropriate potential one
needs to examine the statistics of the topography of the landscape in a way
that has not been done so far. We expect this to be a quite challenging task.

\bigskip

\section*{Acknowledgments}

The work was supported by the Swedish Research Council (VR) and by the EU
Marie Curie Training Site contract: MRTN-CT-2004-512194. The authors are
grateful to Niklas Johansson, Magdalena Larfors and Jian Qiu for useful
comments.

\end{document}